\documentclass[12pt]{iopart}
\usepackage{epsfig}

\begin{document}

\title{Performance of three-photon PET imaging: Monte Carlo simulations.}

\author{Krzysztof Kacperski\dag \ddag \footnote[3]{To
whom correspondence should be addressed
(k.kcperski@nucmed.ucl.ac.uk)} and Nicholas M. Spyrou\dag }

\address{\dag Department of Physics, University of
Surrey, \\ Guildford GU2 7XH, UK}

\address{\ddag Institute of Nuclear Medicine, University College London, \\
Middlesex Hospital, London W1T 3AA, UK}

\begin{abstract}

We have recently introduced the idea of making use of three-photon
positron annihilations in positron emission tomography. In this
paper the basic characteristics of the three-gamma imaging in PET
are studied by means of Monte Carlo simulations and analytical
computations. Two typical configurations of human and small animal
scanners are considered. Three-photon imaging requires high energy
resolution detectors. Parameters currently attainable by CdZnTe
semiconductor detectors, the technology of choice for the future
development of radiation imaging, are assumed. Spatial resolution
is calculated as a function of detector energy resolution and
size, position in the field of view, scanner size, and the
energies of the three gamma annihilation photons. Possible ways to
improve the spatial resolution obtained for nominal parameters:
1.5~cm and 3.2~mm FWHM for human and small animal scanners,
respectively, are indicated. Counting rates of true and random
three-photon events for typical human and small animal scanning
configurations are assessed. A simple formula for minimum size of
lesions detectable in the three-gamma based images is derived.
Depending on the contrast and total number of registered counts,
lesions of a few mm size for human and sub mm for small animal
scanners can be detected.
\end{abstract}



\maketitle

\section{Introduction}
Positron Emission Tomography (PET) is the method of choice in
functional medical imaging, both in clinical practice and research
involving small animals. It uses short-lived positron emitting
radionuclides to mark biologically active substances which can be
traced while being metabolised in the body by detecting co-linear
annihilation photon pairs and reconstructing the image.

Medical imaging is not the only application of positrons. Apart
from basic nuclear and elementary particle research, the physics
of positron annihilation investigates the interactions of
positrons with matter (Charlton and Humberston 2001). They are a
sensitive probe allowing to acquire information about subtle
structural and chemical properties of materials by precise
measurements of annihilation radiation. Three main experimental
methods are employed: positron lifetime spectroscopy, angular
correlation of annihilation radiation and Doppler broadening of
annihilation radiation. Combinations of the above are also in use.
None of these concepts is currently used in PET, apart from
accounting for the non-colinearity of the annihilation photons and
finite positron range contributing to deterioration of spatial
resolution. Annihilation pairs are used merely to determine the
activity distribution changing in time.

We introduced recently the idea of three-photon PET imaging
(Kacperski \etal 2004, Kacperski and Spyrou 2004a). It was shown
that the relatively rare (about 0.5~\% in water) 3-photon
positron annihilations (Ore and Powell 1949, De Benedetti and
Siegel 1954) can also be used for imaging. By detecting the positions
and energies of the three photons, one can easily locate the
\textit{point} where the annihilation occurred. Thus the
amount of information obtained from a single event is higher than
for a 2$\gamma$ pair, where localisation is along a line. The
rate of 3$\gamma$ decay is not only proportional to concentration
of activity but is also sensitive to the local physical and chemical
environments, notably the presence of oxygen (Cooper \etal 1967,
Klobuchar and Karol 1980, Hopkins and Zerda 1990, Kakimoto \etal
1990). This is due to formation of positronium: a positron -
electron bound state, which behaves as an active chemical
particle. 75~\% of the positronium is formed as an ortho-positronium,
triplet state which in vacuum annihilates only into 3$\gamma$ with
a relatively long lifetime of 142~ns. In matter interaction
processes, in general called quenching, that lead to 2$\gamma$
annihilations are usually much faster, reducing the fraction of
three-photon annihilations.

Measuring and imaging oxygenation of tissues is important in
various clinical contexts (Machulla 1999), e.g. in oncology
(tumour hypoxia, radiosensitivity), cardio- and cerebrovascular
disease and infections by anaerobic microorganisms. Intensive
research is ongoing to develop radiotracers for imaging hypoxia.
When the 3$\gamma$ annihilations, which are simply ignored in the
current PET scanners, are recorded, the positron itself, or more
precisely the positronium, could serve as an oxygen-sensitive
tracer. Oxygen is known to be a strong positronium quencher,
therefore hypoxic regions should be characterised by higher
3$\gamma$ rates than those well oxygenated. The oxygenation image
would be obtained alongside e.g. a routine FDG PET, saving cost,
patient radiation dose and inconvenience incurred by extra
specific scan.

In this paper we explore further the idea of 3$\gamma$ imaging by
assessing the dependence of spatial resolution on various system
parameters like the size of the detectors, their energy
resolution, size of the scanner, position within the field of
view as well as the particular combination of three photon energies.
We focus on two scanner configurations commonly used in
practice: a small animal and a full body human scanner. The
expected 3$\gamma$ counting rates are estimated and examples of
images obtained from 3$\gamma$ annihilations are produced. A
formula for the minimum size of detectable lesions with respect to
contrast and number of counts recorded is derived.

\section{Theory}
\label{theory}

\begin{figure}[h]
\begin{center}
\epsfig{figure=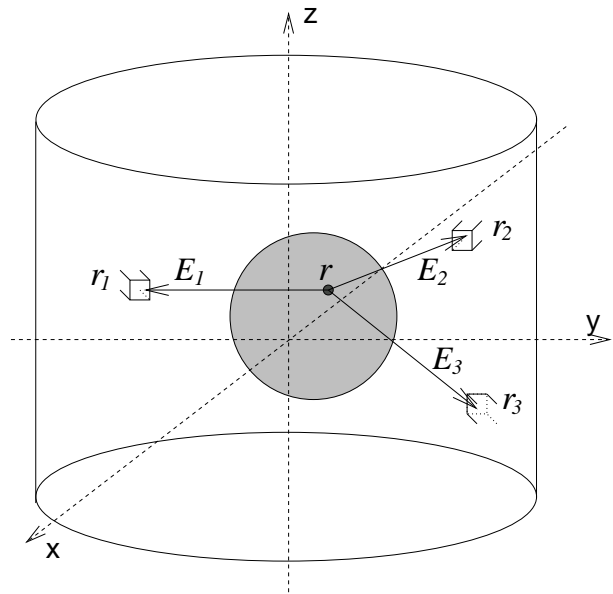,width=8.0cm}
\end{center}
\caption[]{\label{3gscheme} Imaging by three-photon annihilations.
}
\end{figure}

Let us recall the basic principles of 3$\gamma$ PET imaging
(Kacperski \etal 2004). Consider a  $3\gamma$-decay event that
occurs at a point ${\bf r}=(x,y,z)$ (see \fref{3gscheme}).

If the three annihilation photons have energies $E_1$, $E_2$,
$E_3$ and are detected in coincidence by detectors at points ${\bf
r}_1, {\bf r}_2, {\bf r}_3$, respectively, then, from the law of
momentum conservation, we get

\begin{eqnarray}
p_x & = & \frac{E_1}{c} \frac{x-x_1}{|{\bf r} -{\bf r}_1|} +
\frac{E_2}{c} \frac{x-x_2}{|{\bf r}- {\bf r}_2|} + \frac{E_3}{c}
\frac{x-x_3}{|{\bf r} -{\bf r}_3|} =0
\nonumber \\
p_y & = & \frac{E_1}{c} \frac{y-y_1}{|{\bf r}-{\bf r}_1|} +
\frac{E_2}{c} \frac{y-y_2}{|{\bf r}-{\bf r}_2|} + \frac{E_3}{c}
\frac{y-y_3}{|{\bf r}-{\bf r}_3|} =0
\nonumber \\
p_z & = & \frac{E_1}{c} \frac{z-z_1}{|{\bf r} -{\bf r}_1|} +
\frac{E_2}{c} \frac{z-z_2}{|{\bf r} -{\bf r}_2|} + \frac{E_3}{c}
\frac{z-z_3}{|{\bf r} - {\bf r}_3|} =0.
\nonumber \\
\label{momcons}
\end{eqnarray}
In addition, the law of energy conservation has also to be satisfied:
\begin{equation}
E_1 + E_2 + E_3 = m_e c^2 \approx 1022 \mbox{keV}.
\label{encons}
\end{equation}
The energies have a continuous spectrum from 0 to 511~keV with the
probability distribution approximately uniform in the $E_1E_2$
space (Ore and Powell 1949, Chang and Tang 1985). With known
detector positions ${\bf r}_1$, ${\bf r}_2$, and ${\bf r}_3$, the
measurement of photon energies $E_1$, $E_2$, and $E_3$, enables
the solution of the nonlinear set of equations (\ref{momcons}) to
determine the point ${\bf r}$ at which annihilation took place.
Since the energy resolution and size of the detectors are finite,
the location of annihilation is broadened into a region
surrounding the point ${\bf r}$. Note, however, that in contrast
to the $2\gamma$ decay, we obtain (neglecting the finite
measurement accuracy) full information on the position of
annihilation from a {\it single} event, rather than just a line of
response. Unlike the rather intricate image reconstruction methods
required for $2\gamma$ events, one only needs to solve the
nonlinear equations (\ref{momcons}) to retrieve ${\bf r}$ from a
$3\gamma$ event. The image is thus formed as a set of dots, each
corresponding to a single 3$\gamma$ positron decay. It can be
processed further by performing appropriate attenuation correction
and e.g. window averaging to obtain the usual grayscale or
colour-coded pixels display. Let us stress that the recording of
three-photon events is done simultaneously with the prevailing
511~keV photon pairs. 3$\gamma$ imaging is by no means an
alternative to conventional PET, but rather an add-on, making use
of the annihilation radiation which is currently wasted, it
provides extra information. Dividing the 3$\gamma$ image by the
2$\gamma$ one we obtain a map of $3\gamma/2\gamma$ decay
probability ratio. This is actually a new imaging modality.

\section{CdZnTe detectors for PET}

It is clear from the equation (\ref{momcons}) that uncertainty in
the energies $E_1$,  $E_2$, and $E_3$ will result in a spread of
the reconstructed annihilation site ${\bf r}$. As it is indicated
in (Kacperski \etal 2004) and section~\ref{energy} of this paper,
the energy resolution is actually the crucial factor determining
the spatial resolution of $3\gamma$ imaging. With currently
dominating PET scanners which are based on scintillator detectors
(energy resolution typically worse than 15\%) it is not possible
to obtain an accteptable $3\gamma$ image. However, new detector
materials are gradually making their way into PET imaging
technology, notably the room temperature operating semiconductors
CdZnTe (Moses \etal 1994, Scheiber and Gaikos 2001, Verger \etal
2004). The main reason for considering the new material is the
improvement in intrinsic spatial resolution due to precise
depth-of-interaction information, which results in better image
resolution and is of primary importance, specially in small animal
imaging (Stickel and Cherry 2005). Compact scanner design and much
better energy resolution allowing efficient rejection of scattered
events are further advantages. The latter factor also allows
implementation of the 3-photon modality. CdZnTe cameras are
already becoming increasingly popular in SPECT. The price of the
material, processing electronics and implementation of a new
technology are still inhibiting factors, nevertheless, projects to
built prototype small animal CdZnTe PET scanners are already under
way. Many of the material related problems, like slow charge
collection, can be greatly reduced by the use of stacked thin
position sensitive detector arrays (Nemirovsky \etal 2001, Moss
\etal 2001, Redus \etal 2004) with appropriate pulse processing
electronics. Currently energy resolutions of 1\% at 662~keV can be
achieved. Best timing resolutions range between 5 and 8~ns, and
recently even 2.8~ns has been reported (Drezet \etal 2004).

\section{Monte Carlo simulations: small animal and human scanner models}

Performance of PET imaging based on 3-photon annihilations has
been investigated for two kinds of model scanners: small animal
and whole body human. We assumed the usual cylindrical scanner
geometry with square faced detector elements (no dead layers). The
parameters of the scanners are given in Table \ref{skanery}.

\begin{table}[h]
\caption{\label{skanery} Basic parameters of model scanners. }
\begin{indented}
\item[]
\begin{tabular}{@{}lll}
  \br
  Parameter & Small Animal & Human \\
  \mr
  Diameter & 12 cm & 80 cm \\
  FOV & 15 cm & 24 cm \\
  Diameter/FOV ratio & 0.8 & 3.33 \\
  Detector size & 2 mm & 4 mm \\
  Number of detectors in ring & 188 & 628 \\
  Number of rings & 75 & 60 \\
  Detector energy resolution at 662 keV & \multicolumn{2}{c}{1\%} \\
  Minimum detected $3\gamma$ energy ($E_{min}$)  & \multicolumn{2}{c}{150
  keV} \\
Maximum detected $3\gamma$ energy ($E_{max}$)  &
\multicolumn{2}{c}{480
  keV} \\

  \br
\end{tabular}
\end{indented}
\end{table}

In order to investigate their influence on the spatial resolution
chosen parameters have been modified during the simulations, with
all others remaining fixed as in \tref{skanery}.

\subsection{Spatial resolution} \label{sp-res}

For each particular set of parameters at least $10^5$ detected
three-photon events emitted from a point source placed in the
centre of the scanner were simulated. This was assumed to obtain
the point spreads with high statistical accuracy, and does not
reflect the actually achievable counting rates of $3\gamma$
photons; this issue is addressed separately in \sref{scrates}.
Finite size and energy resolution of detectors were the only
sources of blurring in the reconstructed image. We neglected the
range of positrons in matter and residual momentum of the
electron-positron pair. These factors are briefly discussed in
section~\ref{outstanding}. Any photon hitting the surface of a
detector element in the scanner was assumed to be detected at the
centre of that element. The maximum photon position error is
therefore $\sqrt{2}/2$ times the detector size. We assumed that
the energy resolution of the detectors depends on photon energy
according to $\delta E = A \sqrt{E}$, where the constant $A$ was
chosen so that the relative energy resolution at 662 keV is $1\%$
(except section \ref{energy}). After the reconstruction procedure,
as described in \sref{theory}, standard deviation and FWHM of the
reconstructed point were calculated.

\subsubsection{Energy resolution of detectors} \label{energy}

\begin{figure}[h]
\begin{center}
\epsfig{figure=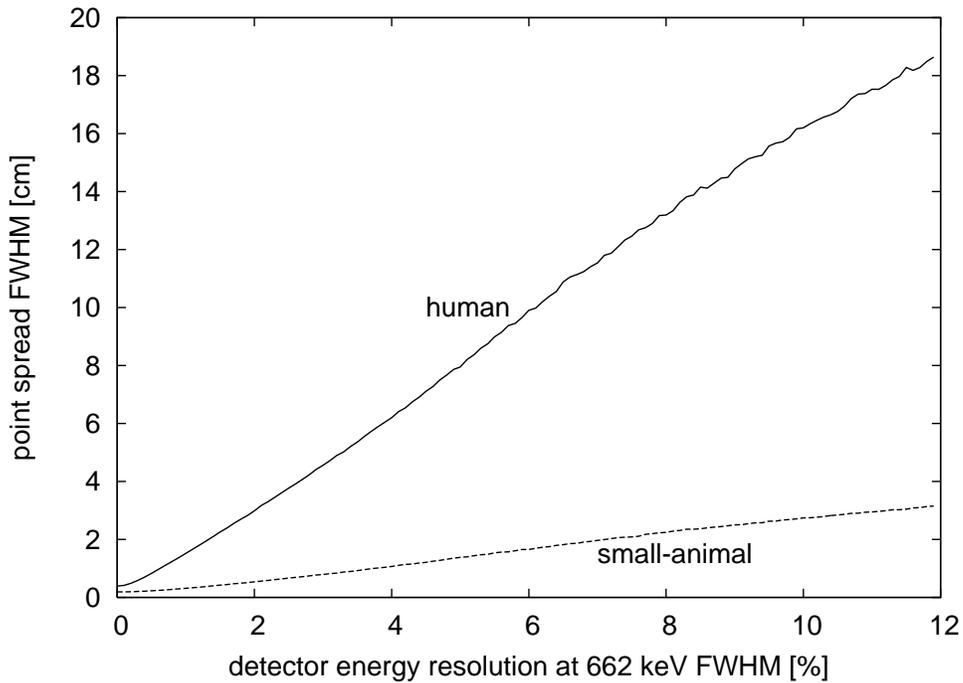,width=9cm,angle=-90}
\end{center}
\caption[]{\label{droddE} Spatial resolution (spread of a point
source) at the centre of the scanner as a function of the relative
energy resolution of the detectors at 662~keV. }
\end{figure}

Let us begin with the dependence of a point source blur on the
energy resolution of the detectors which is shown in \fref{droddE}.
Clearly, good energy resolution of detectors is crucial for
3-photon imaging. With the scintillators currently used for which
the energy resolution is usually worse than $15\%$ at 662~keV one
cannot obtain an acceptable image from the $3\gamma$ events.
However, with CdZnTe detectors of energy resolution $1\%$, or
better, at 662 keV, the spatial resolutions that can be obtained
become acceptable, although worse than typical values for
conventional PET. At very high energy resolutions the influence of
detector size can be noticed (saturation at a non-zero value for
$\delta E =0$), whereas above 1\%, energy resolution is the
dominating source of blurring.

\subsubsection{Position in the FOV}

\begin{figure}[h]
\begin{center}
\epsfig{figure=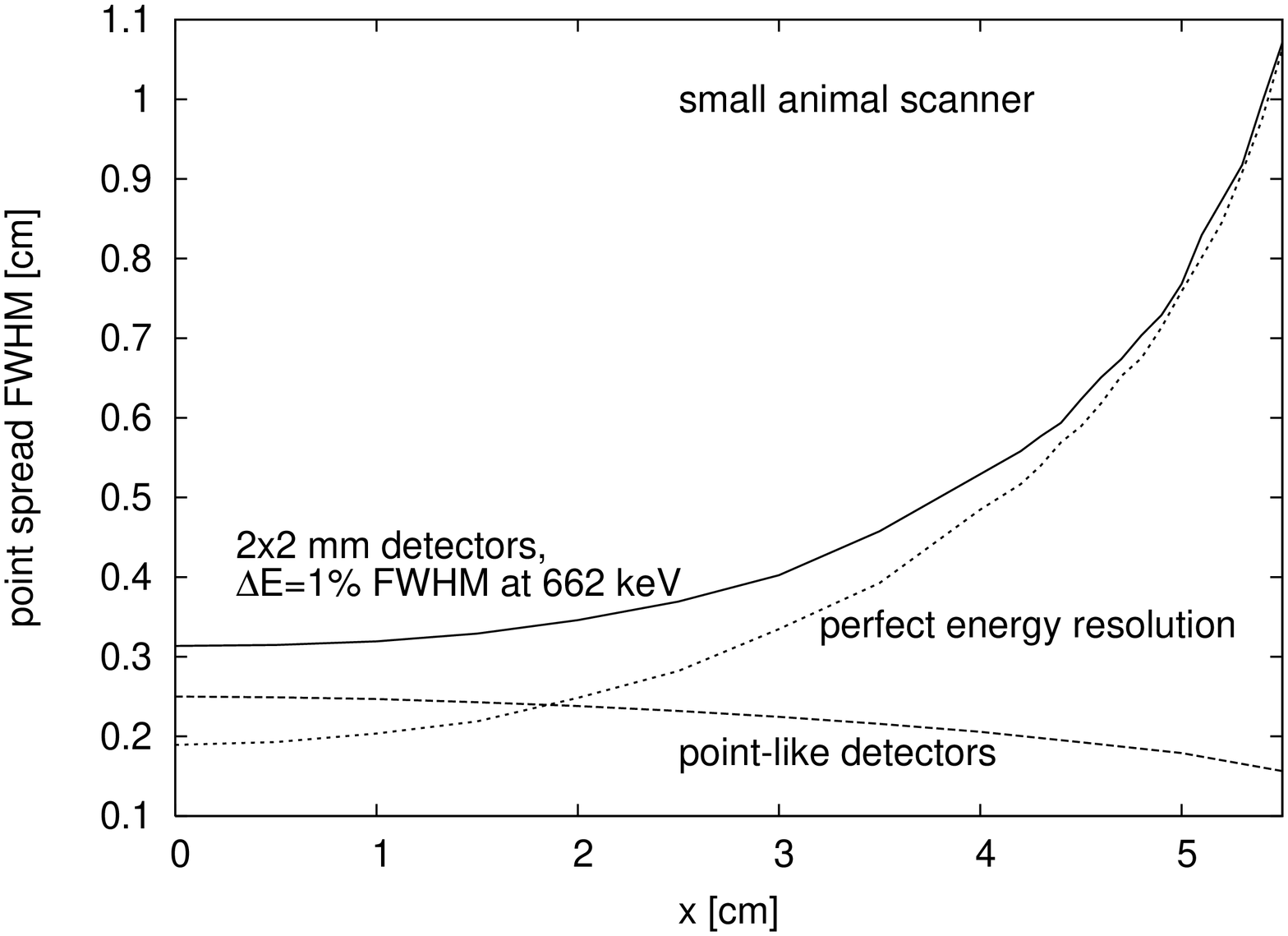,width=9cm,angle=-90}
\epsfig{figure=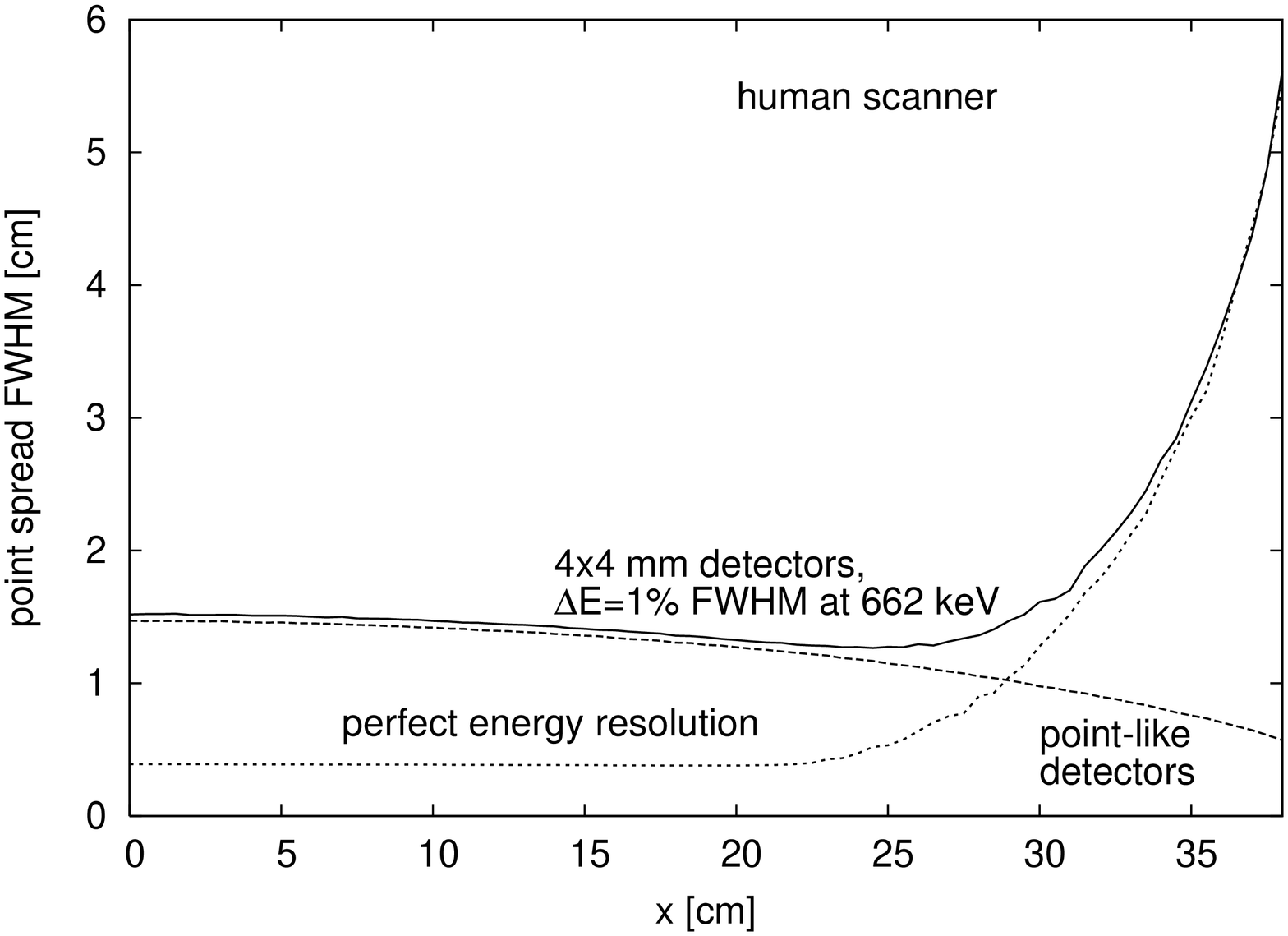,width=9cm,angle=-90}
\end{center}
\caption[]{\label{droddx} Spatial resolution (spread of a point
source) as a function of the transaxial distance from the centre
for the small animal and human scanners. }
\end{figure}

In the following simulations the variation of the spatial
resolution within the field of view has been investigated as the
point source was moved along the radial and axial directions. The
results for the radial direction are shown in figure~\ref{droddx}.

The point spread is a combination of errors in photon energy and
position detection. In order to see the influence of both factors
we ran separate series of simulations assuming perfect energy
detection (error in photon position only), point-like detectors
(error in energy only), and the realistic case when both errors
are present. As one can see in \fref{droddx}, the energy error
contribution increases, while the photon position error decreases
with the transaxial distance from the central axis. In effect, the
resolution is quite uniform across the FOV, deteriorating sharply
only close to the scanner walls. For the human scanner the optimal
resolution is actually achieved at about 25 cm off-centre.

\begin{figure}[h]
\begin{center}
\epsfig{figure=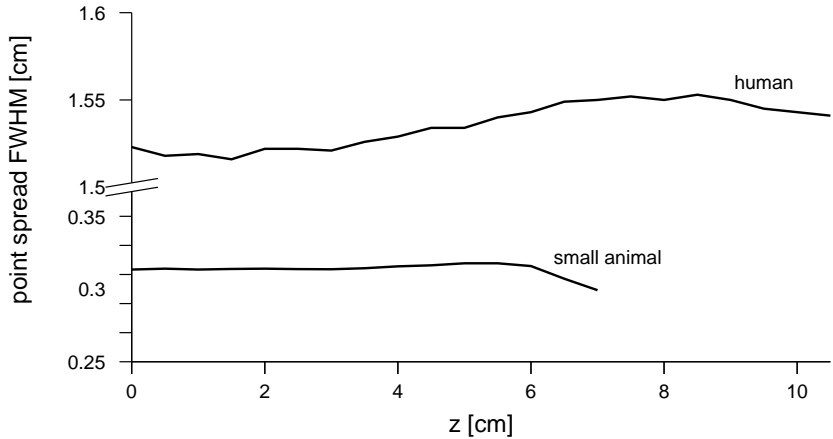,width=12.0cm}
\end{center}
\caption[]{\label{droddz} Spatial resolution (spread of a point
source) as a function of the distance from the centre along the
$z$ axis of the scanners. }
\end{figure}

The variability along the $z$ axis (\fref{droddz}) is even
smaller, remaining below $2.5 \%$ and $4 \%$ for human and small
animal scanner, respectively. The results of other simulations
obtained for the point at the centres of scanners are therefore
representative for the entire useful FOV.

It should also be noted that the shape of the point spread
function is spatially nonuniform. For the point at the centre the
spread in the transaxial direction equals 1.06 cm and 0.2 cm FWHM,
respectively, for human and small animal scanner, whilst in the $z$
direction it is 0.26 cm and 0.136 cm, respectively. As we move off from
the central axis the PSF becomes elongated in the radial
direction.

\subsubsection{Size of detectors}

\begin{figure}[h]
\begin{center}
\epsfig{figure=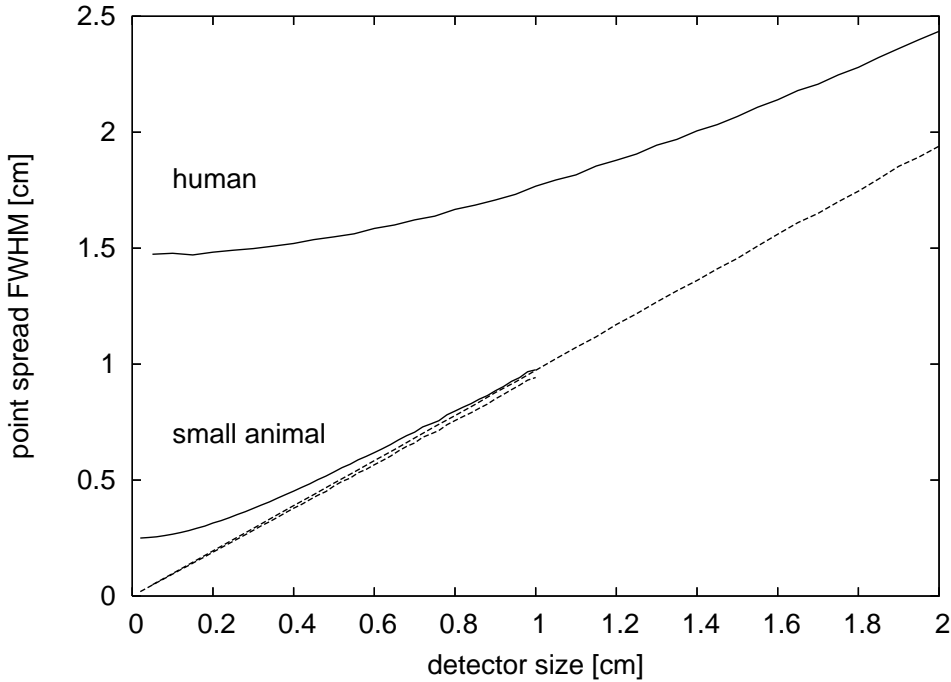,width=9cm,angle=-90}
\end{center}
\caption[]{\label{drodda} Spatial resolution (spread of a point
source) as a function of detector size. The dashed lines show
the same dependance assuming perfect energy resolution (no energy
blur). }
\end{figure}

Keeping the geometric proportions of the scanners fixed as in
table~\ref{skanery} the size of (square) detectors was changed,
and their number adjusted accordingly to keep the sizes of scanners
constant. \Fref{drodda} shows the results of the simulations with
detector energy resolution $1 \%$ at 662 keV (solid lines) and
perfect energy detection (dashed lines).

The overall dependence, similar to \fref{droddE}, is close to
linear. Comparison of both curves for each scanner reveals the
contributions of the two sources of error and indicates the
direction of potential improvements. For our model human scanner
with detector size 4~mm the reconstructed point spread results
mainly from the energy detection error (cf. also \fref{drodxx}),
therefore decreasing the size of detectors, without improving
their energy resolution, would hardly improve the spatial
resolution. For the small animal scanner with 2~mm detector
elements the contribution from photon position detection error is
higher, so reducing effective detector size would bring some
performance improvement, although energy resolution is still more
critical. The influence of detector size becomes more important
for points far from the centres of scanners (see \fref{drodxx}).

\subsubsection{Size of scanner}

\begin{figure}[h]
\begin{center}
\epsfig{figure=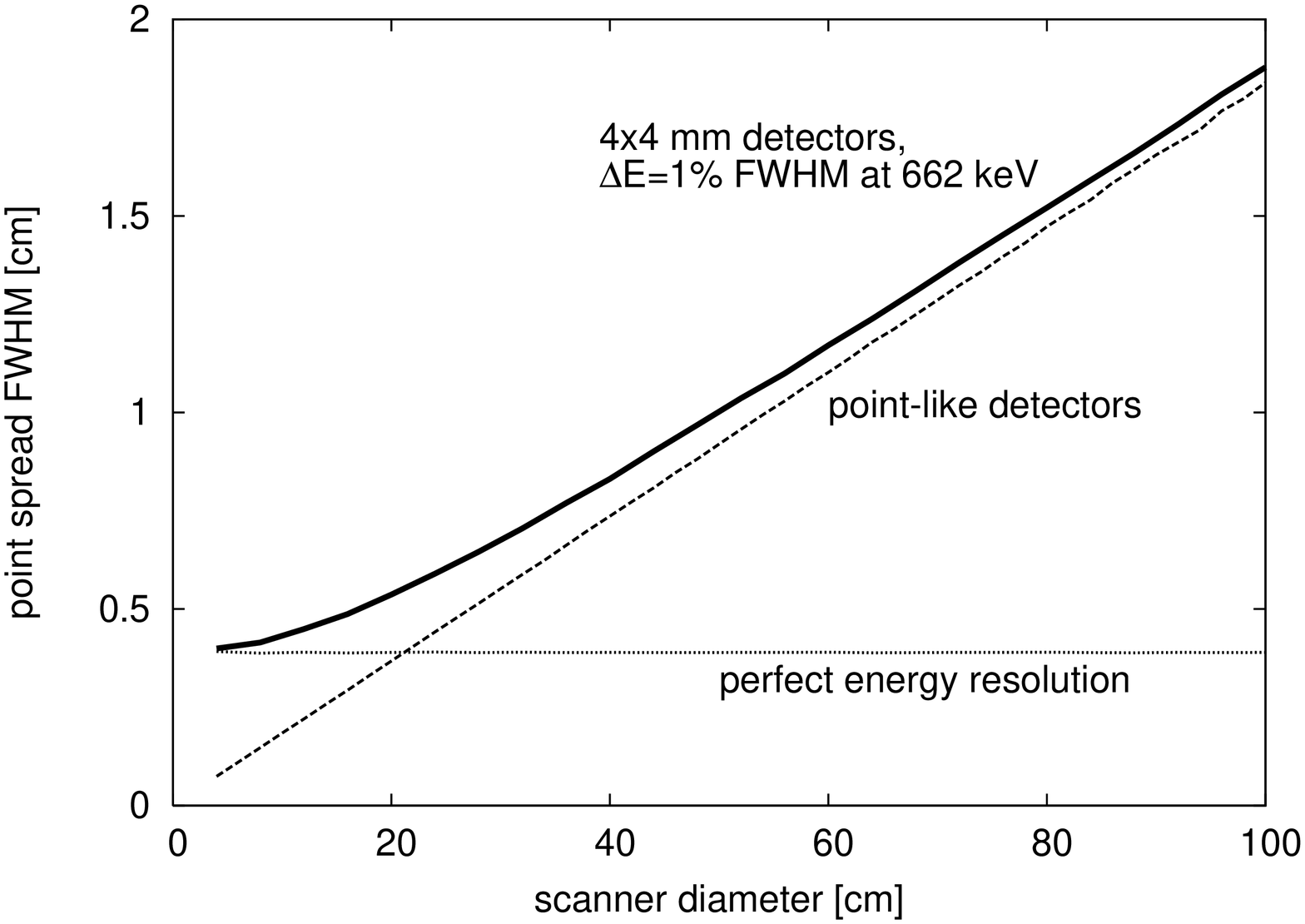,width=9cm,angle=-90}
\epsfig{figure=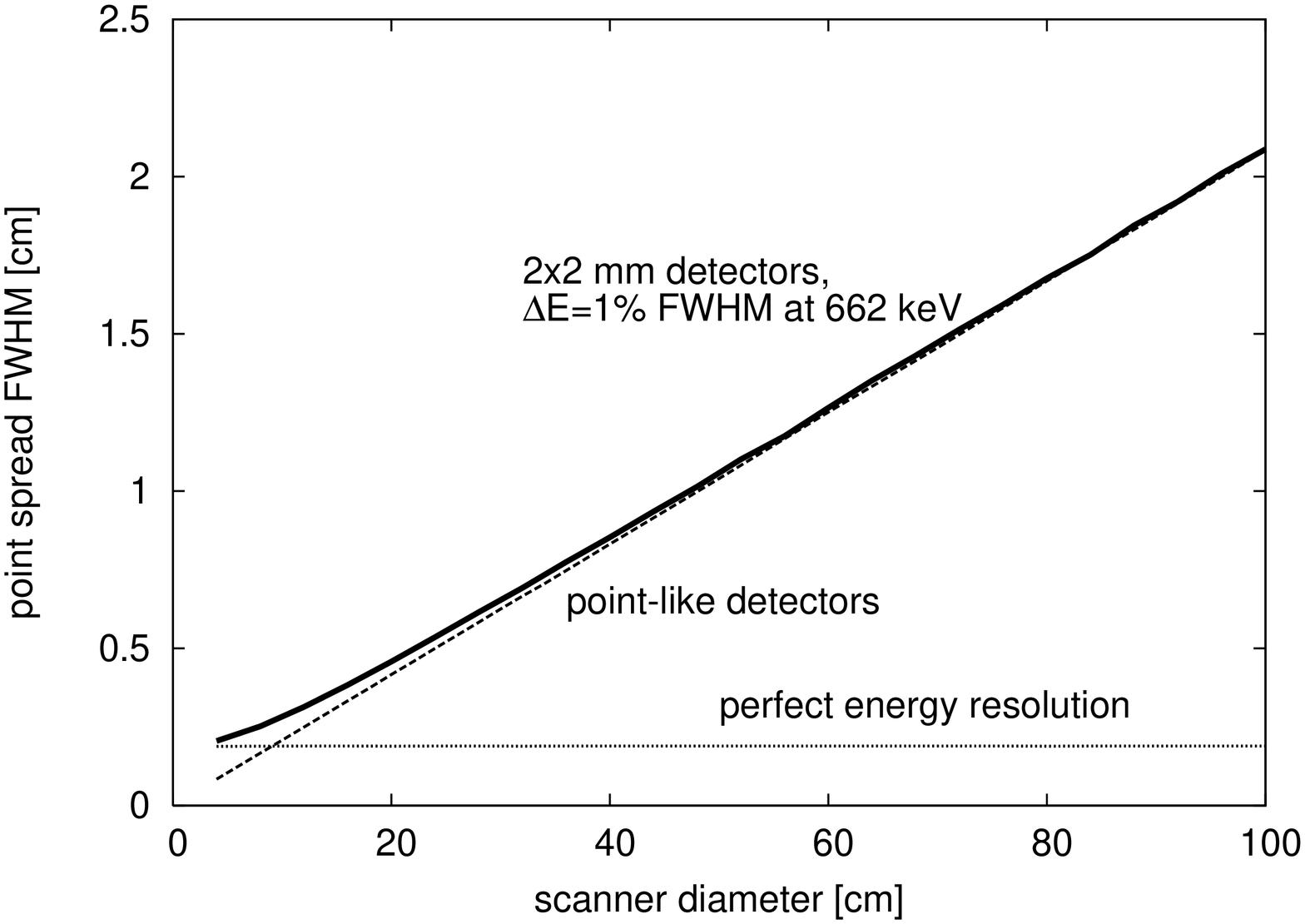,width=9cm,angle=-90}
\end{center}
\caption[]{\label{drodR} Spatial resolution (spread of a point
source) as a function of scanner size $D$ for human (upper plot)
and small animal (lower plot) scanner proportions. Solid lines: finite photon
energy and position resolution, dashed: point size detectors,
dotted: perfect energy resolution (no energy blur). }
\end{figure}

The influence of the scanner size on the spatial resolution was
studied by changing the diameter $D$ of the scanners and their
axial length (number of rings) to preserve the geometric
proportions as in table~\ref{skanery}. The size of detectors was
kept constant and their number adjusted to cover the whole
cylinder. Again, three separate simulations were run for each
scanner: assuming perfect detection of photon energy, or position,
and modelling errors in both. The results can be seen in
\fref{drodR}.

One can clearly see that the near-linear dependence is due to
the energy resolution error which depends linearly on scanner
size. The detector size component remains constant. For both human
and small animal scanners reducing scanner size as much as
possible is desirable from the point of view of $3\gamma$ imaging.
This dependence is the main reason why the spatial resolution of
the small animal scanner is significantly better than that of
human the scanner.

\subsubsection{Energies of photons and cut-off energies}
\label{sec-e1e2}

\begin{figure}[h]
\begin{center}
\epsfig{figure=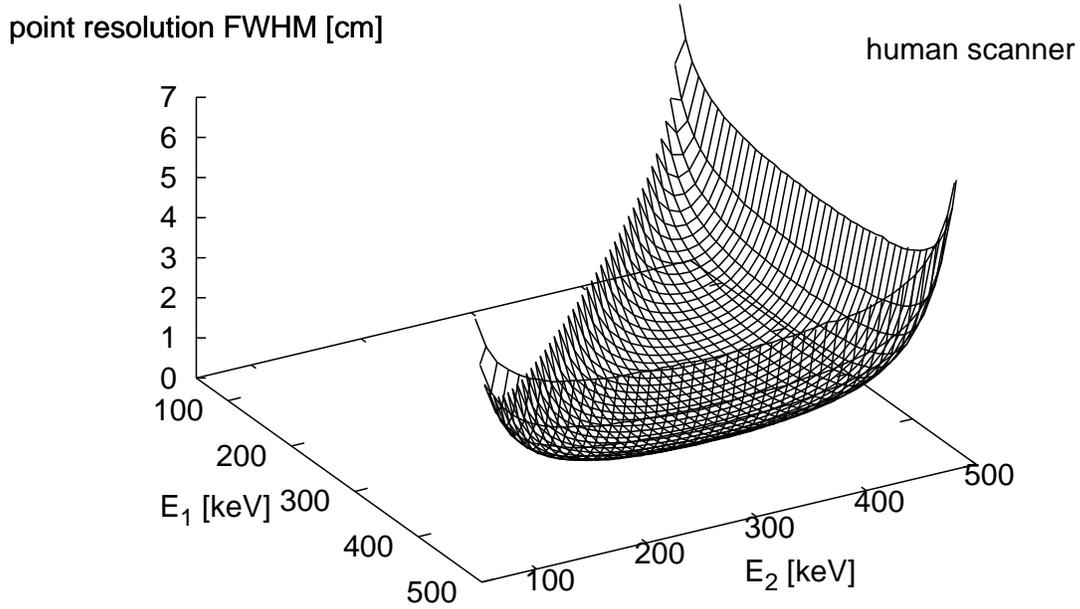,width=8cm,angle=-90}
\end{center}
\caption[]{\label{drodE1E1} Spatial resolution (spread of a point
source) as a function of photon energies for the human scanner.}
\end{figure}

As it was mentioned in \sref{theory}, the three photons can have
any combination of energies satisfying (\ref{encons}) and
(\ref{momcons}). The error of the annihilation site $r$ recovered
from (\ref{momcons}) is a rather complex function of $E_1$, $E_2$
and $E_3$. We have calculated the FWHM spreads of a point source
at the centre of the human scanner emitting photons with a
particular energy combination satisfying (\ref{encons}) and
(\ref{momcons}) (\fref{drodE1E1}). The spreads were averaged over
the remaining free parameters (directions of emission).

\begin{figure}[h]
\begin{center}
\epsfig{figure=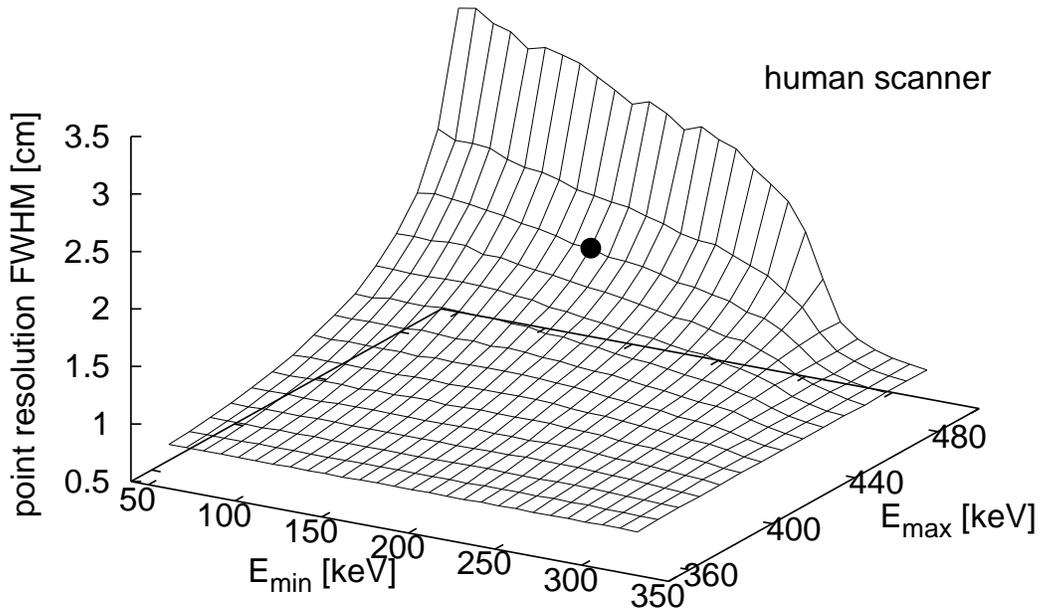,width=8cm,angle=-90}
\end{center}
\caption[]{\label{drodEminEmax} Spatial resolution (spread of a
point source) as a function of upper and lower energy threshold
for $3\gamma$ photons in human scanner. The dot indicates the
energy window used in all simulations of this paper (cf.
\tref{skanery})}
\end{figure}

The resolution is optimal for the symmetric three-photon decay
with all energies equal 340.7~keV and deteriorates as we move
towards extreme values. In practice we never detect energies from
the entire $3\gamma$ spectrum 0-511~keV. At the upper end the
511~keV photopeak resulting from the dominating $2\gamma$ decays has to be
cut-off. On the other hand, there is a detector and noise related
low energy detection limit. \Fref{drodE1E1} indicates yet another
reason to avoid registering $3\gamma$ decays with extreme
energies: they introduce large errors in position reconstruction.
In \fref{drodEminEmax} the point spread of a point
source as a function of upper and lower $3\gamma$ energy
thresholds was plotted.
The smaller window around the optimal 340.7 keV we set, the better
spatial resolution we achieve. There is, however, the common
sensitivity-resolution trade-off: as the window narrows, the number
of counts detected drops, so a compromise choice has to be made.

\Fref{drodE1E1} also suggests a relatively easy way of
improving spatial resolution without losing counts. In all our
simulations we treat all the reconstructed $3\gamma$ annihilation
points equally. We could, however, using the relation shown in
\fref{drodE1E1}, assign a weight to each point according to the
particular energy combination, so that counts with energies close
to optimal ($E_1=E_2=E_3=340.7$ keV) would contribute more than
those with energies close to extreme. Such approach would probably
improve spatial resolution by about 30~-~40~\%.

\subsection{Counting rates and random scattered $3\gamma$ events}
\label{scrates}

Poissonian noise is one of the key limiting factors in nuclear
medicine imaging. With the probability of three-photon
annihilations being two to three orders of magnitude lower than that of
two-photon, the question of sufficient number of counts obviously
needs to be addressed. Will it be enough to obtain meaningful
images?

We will assess the rates of $3\gamma$ counts for our two model
scanners with phantoms in the form of spheres filled with water
and uniform positron activity distribution. The diameters of the
spheres are 20 cm and 4 cm for the human and small animal scanners,
respectively; they are placed at the centres of the scanners. The
counting rate $C_{3\gamma}$ of true $3\gamma$ decays is in general
determined by probabilities of emission, detection and
attenuation. It can be expressed by the formula
\begin{equation}\label{3grate}
  C_{3\gamma} =A\: r_{3\gamma/2\gamma} \: E_{cut} \: \eta \: d_{eff} \: g_{3\gamma} \:
  (\mbox{e}^{-A \tau} + q_1 A \tau \mbox{e}^{-A \tau} + q_2 (A \tau)^2 \mbox{e}^{-A \tau}),
\end{equation}
where $A$ is the total activity in the phantom. The meaning of the
factors and their values for the two model systems is summarized
in \tref{cr-param}. The expression in brackets approximates the
probability that a true three-photon decay will not be obscured by
(one or more) coincident $2\gamma$ events. The three terms
correspond to zero, one and two $2\gamma$ pairs respectively.
Higher order coincidences are negligible at the activities
relevant for PET. The factors $q_1$ and $q_2$ are the
probabilities of unique identification of the true $3\gamma$ event
despite the coincident $2\gamma$ photons. This is possible because
the coincident photons may just miss the scanner, or, if they are
detected, their energies are quite different from those of
$3\gamma$ annihilation.
\begin{table}[h]
\caption{\label{cr-param} Parameters influencing $3\gamma$
counting rate (cf. \eref{3grate}). }
\begin{indented}
\item[]
\begin{tabular}{@{}llll}
  \br
Symbol &  Parameter & Small Animal & Human \\
  \mr
  $\eta$ & attenuation factor & 0.65 & 0.11 \\
  $g_{3\gamma}$ & geometry factor & 0.51 & 0.052  \vspace{0.2cm} \\

  $q_1$     &  probability that one $2\gamma$ pair &   0.94   &  0.98 \\
            &  will not interfere with $3\gamma$ detection &  &   \vspace{0.2cm} \\

  $q_2$    &  probability that two $2\gamma$ pairs &  0.5   &  0.8 \\
           &   will not interfere with $3\gamma$ detection &  &  \vspace{0.2cm}\\

  $r_{3\gamma/2\gamma}$ & $3\gamma/2\gamma$ emission probability ratio &  \multicolumn{2}{c}{0.004}  \vspace{0.2cm} \\
  $E_{cut}$ & fraction of $3\gamma$ photons in the  &  \multicolumn{2}{c}{0.655} \\
            &  ($E_{min}$, $E_{max}$) window  &  &   \vspace{0.2cm} \\
   $d_{eff}$ & detection efficiency & \multicolumn{2}{c}{ $0.85^3$ = 0.614} \\

  \br
\end{tabular}
\end{indented}
\end{table}
Numerical values of the parameters were obtained from simulations
and analytical calculations. We assumed the scanners to be made of
3~cm thick CdZnTe detectors, which yields average full energy photopeak
efficiency for the $3\gamma$ photons of about $85\%$. The
$3\gamma/2\gamma$ ratio was derived from positron lifetime
experimental data for water and organic liquids. In fact, it
depends sensitively on the chemical composition of the solution.
Usually dissolved ions as well as gases (in particular oxygen)
lead to a decrease of $3\gamma$ rate. On the other hand, the ratio
increases in liquids of smaller surface tension. The
$r_{3\gamma/2\gamma}$ ratio for pure, degassed water is about
0.5\%; in alcohols it reaches about 0.7\%. The minimum value
occurring in substances (e.g. metals) where no positronium is
formed is 0.27\%. Our choice of $r_{3\gamma/2\gamma} = 0.4\%$
seems therefore reasonable and not too optimistic.

\begin{figure}[h]
\begin{center}
\epsfig{figure=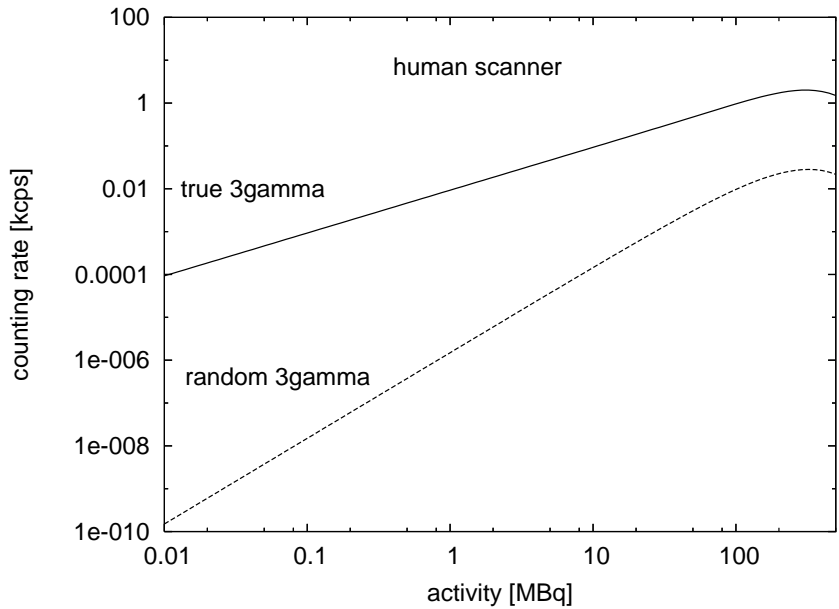,width=8cm,angle=-90}
\end{center}
\caption[]{\label{crates-h} Estimated true and random $3\gamma$
counting rates for human scanner.}
\end{figure}

\begin{figure}[h]
\begin{center}
\epsfig{figure=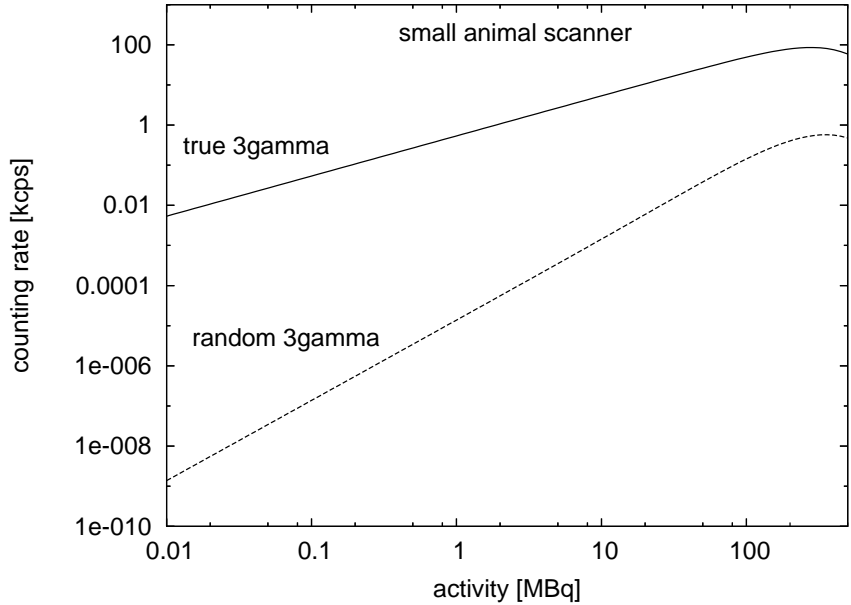,width=8cm,angle=-90}
\end{center}
\caption[]{\label{crates-a} Estimated true and random $3\gamma$
counting rates for small animal scanner.}
\end{figure}

The $3\gamma$ counting rates for both scanners are plotted in
figures \ref{crates-h} and \ref{crates-a}. They can amount up to
about 1000~cps for human scanner and 80,000~cps for small animal
scanner. The difference is mainly due to the larger solid angle
covered and much smaller attenuation for the small object.

In conventional $2\gamma$ PET, apart from the true coincidence
counts forming the image, there are scattered and randomly
coincident photons contributing to image noise. In the case of
$3\gamma$ imaging those two kinds of events have to be considered
simultaneously. Because of the resolution requirement we need to
use high energy resolution detectors (see section~\ref{energy}).
The unique properties of $3\gamma$ decay, in particular
the energy condition (\ref{encons}) enables then to efficiently
distinguish the true events from accidental coincidences.
Nevertheless, it can happen that all the conditions are fulfilled
by chance leading to spurious counts. Let us estimate the rates of
such counts for our model scanners. Because of the small $3\gamma$
decay rate the probability of two or more $3\gamma$ events
occurring and being detected within the resolving time is very
small. The main source of spurious counts are the randomly
coincident $2\gamma$ annihilation photons that are scattered or
partially detected, so that they fall within the energy range of
the $3\gamma$ spectrum. Let us denote by $P(E|511)$ the
probability of a 511 keV photon emitted from the phantom to
deposit the energy $E$ in a detector. It is actually a convolution
of the spectrum of photons leaving the phantom (primary and
scattered) and the detector response function. The probability of
three coincident photons randomly fulfilling condition
(\ref{encons}) within the resolution determined window $\Delta E$
can be approximately written as
\begin{equation}
I_1 = \int_{W(\Delta E)} P(E_1|511) P(E_2|511) P(E_3|511) dE_1
dE_2 dE_3 \label{I1}
\end{equation}
where the integration is over the region $W(\Delta E)$ between the
planes $E_1 + E_2 +E_3 = 1022$~keV~$\pm~\Delta E$ in the
$E_1E_2E_3$ space, further limited by $E_{min}$ and $E_{max}$, the
upper and lower limits of detected $3\gamma$ spectrum. Note that
$I_1$ does not include any geometry dependence. In our estimation
$P(E|511)$ in the interval ($E_{min}$, $E_{max}$) can be
approximated by a linear function with parameters obtained from
Monte Carlo simulations. The rate of false $3\gamma$ counts can be
in general expressed as
\begin{equation}
C_{f3\gamma} = \sum_{n=1}^{\infty} \frac{(A \tau)^n \mbox{e}^{-A
\tau}}{n!} f(n) \label{false3g1}
\end{equation}
The fraction is the Poissonian probability of $n+1$ decays
occurring within the resolving time $\tau$, and $f(n)$ is the
probability that at least one triplet of photons will be detected
and falsely accepted as true $3\gamma$ event. Practically, it is
sufficient to consider just the first two terms in
(\ref{false3g1}), and then the false $3\gamma$ counting rate
becomes
\begin{flushleft}
\begin{equation}
\begin{array}{ll}
C_{f3\gamma} =  \mbox{e}^{-A \tau} & \left\{    A \tau \left[
g_1^3 (1-g_1) +
\left(%
\begin{array}{c}
  4 \\
  3 \\
\end{array}%
\right) g_1^4 \right] \right. \\
& +  \frac{(A \tau)^2}{2} \left[\left(
\begin{array}{c}
  6 \\
  3 \\
\end{array} \right)g_1^6
 +
\left(
\begin{array}{c}
  5 \\
  3 \\
\end{array} \right)g_1^5 (1-g_1) \right. \\
& \left. \left. + \left(
\begin{array}{c}
  4 \\
  3 \\
\end{array} \right)g_1^4 (1-g_1)^2
 + g_1^3 (1-g_1)^3 \right]  \right\}  I_1 P_{obj}
\end{array} \label{false3g2}
\end{equation}
\end{flushleft}
The two terms in the curly brackets correspond to two and three
two-photon annihilations occurring within the resolving time.
$g_1$ is the probability to hit the scanner by a single photon
(geometric factor). We assumed that the directions of photons are
uncorrelated, which is not quite true, but works well as an
approximation. Even when three detected accidental photons fulfil
condition (\ref{encons}) $\pm \Delta E$, when inserted into
(\ref{momcons}) most of them would either give no real solution,
or generate a point outside the object. $P_{obj}$ is the fraction
of events yielding a point within the object, contributing to
$3\gamma$ image noise. For our scanners and phantoms
$P_{obj}=0.02$, the single photon geometric factors $g_1$=0.287
for the human and $g_1$=0.781 for the small animal scanner.

The rates of random $3\gamma$ events (figures \ref{crates-h} and
\ref{crates-a}) remain very low compared to true $3\gamma$ even at
high activities, so they do not contribute significantly to image
noise. This signal-to-noise ratio is, however, sensitive to the
detection efficiency of the detectors and may worsen significantly
for photopeak efficiencies lower than those assumed.

\subsection{Examples of images and detectability of lesions}

\begin{figure}[h]
\begin{center}
\epsfig{figure=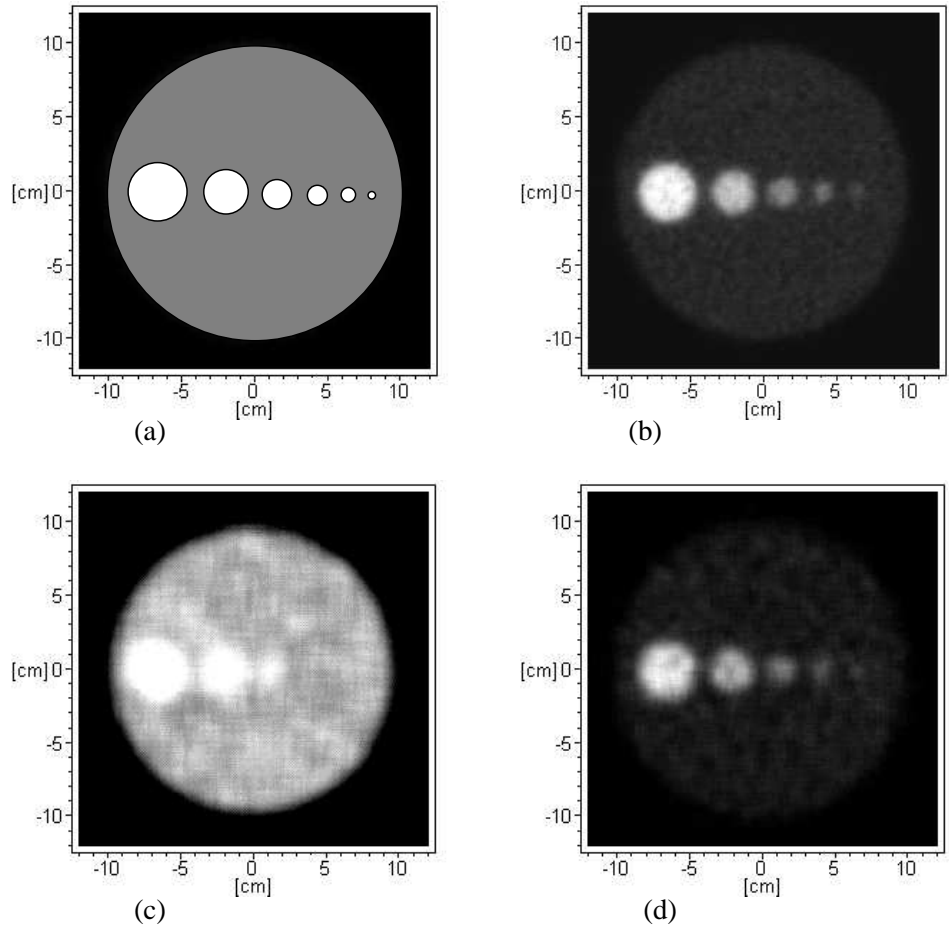,width=13.0cm}
\end{center}
\caption[]{\label{images} Examples of images obtained from
$3\gamma$ events in simulation of the human scanner. The water
filled phantom (a) consisted of 6 high activity spheres of
diameters 4, 3, 2, 1.5, 1 and 0.5 cm embedded in the 20 cm sphere
with background activity, which was placed at the centre of the
human scanner (cf. Table~\ref{skanery}). The images are mean
filtered projections of 1 cm thick central slice. (b): background
count density $n_b$=240 counts/cm$^3$, contrast $c$=3, mean filter
kernel (square averaging window) size: 5~mm  (c):
$n_b$=240~counts/cm$^3$, $c$=0.2, filter 17~mm, (d): $n_b$=30
counts/cm$^3$, $c$=3, filter 9~mm.}
\end{figure}

In order to generate examples of images we used the human scanner
with the phantom described in the previous section and added
lesions in the form of spheres of variable size, lined up across
the centre of the phantom as shown in \fref{images}~(a). We define
the lesion contrast as $c=(n_l-n_b)/n_b$, where $n_l$ is the
density of registered $3\gamma$ counts in the lesion, and $n_b$
that in the background. In \fref{images} a few examples of
$3\gamma$ images are shown.
Image (b) corresponds to the maximum $3\gamma$ counting rate for
the human scanner of about 1000~cps (cf. \fref{crates-h}) scanned
for about 20 min. At contrast $c=3$ the 1~cm lesion is clearly
visible, however the one of 0.5~cm diameter is not. When contrast
drops to $c=0.2$ (c), the detectability of lesions deteriorates
significantly. On the other hand the 1~cm lesion is still visible
even if the count statistics is reduced by almost order of
magnitude (d).

\begin{figure}[h]
\begin{center}
\epsfig{figure=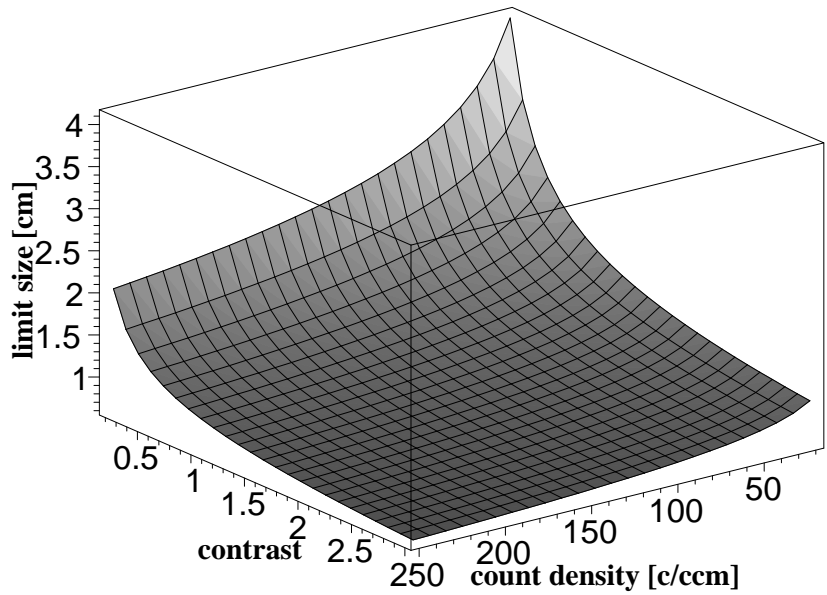,width=12.0cm}
\end{center}
\caption[]{\label{visib} Minimum size of spherical lesion to be
distinguishable against the background as a function of contrast
and density of counts in the background for a scanner of spatial
resolution 1.5 cm FWHM (human).}
\end{figure}

Detectability of lesions clearly depends on their size and
contrast. Let us examine this dependence to assess the limits of
three-photon imaging. In order for a lesion to be distinguishable
against the background its total number of points (counts) above
the background level, reduced by its possible statistical
fluctuations, should be significantly higher than the average
statistical fluctuation of the number of points in a background
region of the same size as the lesion. This can be expressed as
\begin{equation}
V_0 n_b c - \xi_1\sqrt{V_0 n_b (c+1)} = \xi \sqrt{V n_b}.
\label{viseq1}
\end{equation}
$V_0$ is the original volume of the lesion without the partial
volume effect. In the image, however, the blurring due to finite
resolution of the scanner has to be taken into account; the
blurred lesion volume is denoted by $V$. The parameters $\xi$ and
$\xi_1$ specify what ``significantly higher'' actually mean, in
other words they control the confidence level of our assessment.
Here we assume $\xi$ = $\xi_1$ = 2, which corresponds roughly to
95\% confidence level. For a spherical lesion
equation~(\ref{viseq1}) takes the form
\begin{equation}
\sqrt{\frac{\pi}{6} n_b} \; c \: {a_0}^3 - \xi_1\sqrt{(c+1)
{a_0}^3} = \xi \sqrt{(a_0+\sigma)^3} \label{viseq2}
\end{equation}
where $a_0$ is the original diameter of the lesion and $\sigma$ is
the FWHM spatial resolution of the scanner.
Equation~(\ref{viseq2}) is polynomial and can be solved
numerically for $a_0$. In the absence of scanner blurring
($\sigma=0$) it can be reduced to the formula
\begin{equation}
a_0 = \sqrt[3]{\frac{6(\xi+\xi_1 \sqrt{c-1})^2}{\pi n_b c^2}}
\label{a0}
\end{equation}
The numerical solution of \eref{viseq2} for our model human
scanner ($\sigma=1.5$~cm) is plotted in \fref{visib}.

The formula (\ref{viseq2}) works quite well for the simulated
images. The calculated minimal detectable sizes of lesions for the
images in \fref{images} are: (b)~6.2~mm, (c)~20.8~mm, (d)~10~mm.
For the model small animal scanner (spatial resolution
0.32~cm~FWHM) with 2~cm water phantom (\sref{scrates}) of activity
50~MBq the $3\gamma$ counting rate can reach over 50~kcps (see
\fref{crates-a}). With the much smaller phantom the count density
is about $1.5\cdot10^6$~c/cm$^3$ in a 20 min. scan. Then
sub-millimeter detectability is achieved at contrasts $>$ 0.7. At
$c=3$ lesions as small as 0.6~mm should be visible.

The formula (\ref{viseq2}) allows to fine tune the trade-off
between the resolution and number of counts represented by the
relation shown in \fref{drodEminEmax}. Depending on the contrast
of lesions, it may be of advantage to narrow the $3\gamma$ window,
losing some of the counts (decrease $n_b$), but improving the
spatial resolution (decrease $\sigma$) to achieve the best
detectability.

\subsection{Attenuation}

\begin{figure}[h]
\begin{center}
\epsfig{figure=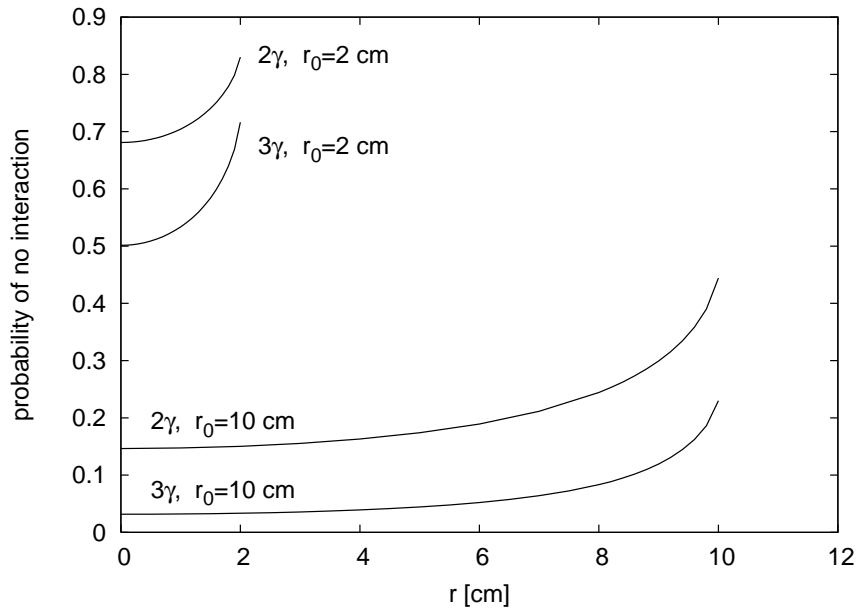,width=8.0cm,angle=-90}
\end{center}
\caption[]{\label{atten} Average probability of emission from a
sphere filled with water (diameters 4 cm and 20 cm) without
interaction for a $2\gamma$ annihilation pair and the three
photons from a $3\gamma$ decay.}
\end{figure}

In $3\gamma$ imaging we unfortunately do not have the advantage of
relatively easy attenuation correction as in conventional
$2\gamma$ PET, where the attenuation along any line of response is
constant, independent of the site of annihilation and can be
measured directly by a transmission scan. For each point in an
object and each particular combination of energies and emission
directions of the three photons the attenuation factor will be
different. The problem of attenuation correction is therefore
rather complex, similar to for example scatter correction in
conventional PET or SPECT. Using Monte Carlo simulations and
having the map of attenuation coefficients for the object one can
compute a map of attenuation factors for each point of the object
averaged over the free parameters of $3\gamma$ emission. The
results for our water phantoms are shown in \fref{atten}.

In general attenuation for $3\gamma$ photons is higher than for
$2\gamma$ because of lower energies and longer effective paths. It
is also strongly nonuniform, being significantly higher for points
deep inside the body.

\section{Outstanding questions and directions of future studies}
\label{outstanding}

In this paper we concentrated on the image quality and performance
of the new $3\gamma$ PET modality, leaving for future studies the
question of biological and clinical significance of the
information extracted from the three-photon annihilations. Precise
measurements of $3\gamma$ yields in biological samples are
necessary to determine its variability in a living organism, as
well as sensitivity to parameters like the level of oxygen. Our
results will allow to assess whether the new information can be
extracted by PET systems based for example on CdZnTe detectors.

In our simulations we neglected the effect of finite positron
range and non-zero residual momentum of the annihilating
positron-electron pair. Both effects add extra blurring of
magnitude similar to that for the two-photon case. That is in most
cases negligible in comparison to the effects of detector size and
energy resolution.

In our estimation of counting rates we did not consider the
contribution of activity outside the field of view, which is
important in conventional clinical PET. It would add extra random
$3\gamma$ counts. However, our calculations show (\sref{scrates})
that for the assumed scanner parameters the rate of random
$3\gamma$ events is at least 3 orders of magnitude lower than that
of true $3\gamma$. Extra activity contributing to randoms would
mean a shift on the randoms curves in figures~\ref{crates-h} and
\ref{crates-a} along the activity axis typically by a factor of
2-4. It means that the random $3\gamma$ rate would still remain
very small compared to the true $3\gamma$ and would not pose a
real problem. The above is true for pure positron emitters, which
are most commonly used in PET. If there are nuclear gamma photons
accompanying positron emissions, the rate of triple coincidences
and false $3\gamma$ events may increase significantly, although
still most of them could be rejected due to the energy
conservation condition (\ref{momcons}).

The limitations of $3\gamma$ imaging seem to be in the first place
the low number of counts, especially with high attenuation (large
patients, areas located deep in the body), and the spatial
resolution limited mainly by the energy resolution of detectors.
While one can not realistically do much to increase the counting
rates achieved in our model scanners (\sref{scrates}), the spatial
resolution leaves plenty of room for improvement. One way would be
to rectify the energy resolution of the detectors (see
\sref{energy}). Although the technology of room temperature
semiconductor detectors is making constant progress, we probably
can not expect a significant further improvement in energy
resolution compared to that assumed in our simulations. An
alternative is to use e.g. HP-Ge detectors offering resolutions of
the order of 0.3~\%. However, the other properties, and in
particular the need for cryogenic cooling and cost make the choice
rather impractical at least for clinical PET. Their use for a
dedicated small animal PET system is, however, not excluded
(Philips \etal 2002). A much more feasible way to improve the
resolution of $3\gamma$ imaging is through the reconstruction
process. In this paper we used the simplest possible procedure by
solving equations (\ref{momcons}) for each registered $3\gamma$
event and producing a ``set of points'' image. The easiest way to
improve resolution is by making use of the energy dependence of
the positioning error (\fref{drodE1E1}) as indicated in
\sref{sec-e1e2}. Another possibility is to account for the unique
non-symmetric shape of the point spread function which can be
calculated for each combination of detected $3\gamma$ photons. It
could be incorporated in the framework of a statistical iterative
reconstruction method like ML-EM (Shepp and Vardi 1982), known to
produce superior quality images, however, for the price of
significant complexity and computation time.

\section{Conclusions}

We have studied thoroughly the main characteristics of
three-photon imaging in positron emission tomography. Clearly, it
is one of the directions for future development of this rapidly
expanding imaging technique. The main prerequisite for the new
modality to be implemented is a scanner based on high energy
resolution detectors, like CdZnTe. This is no longer a futuristic
dream. Several groups and manufacturers' laboratories are working
on such devices, and probably within the next few years we will
see first prototypes in practical use.

Our simulations show that for typical scanner configurations, with
currently available CdZnTe detector properties, good quality
$3\gamma$ annihilation images can be obtained. They do not match
those from conventional PET in terms of spatial resolution and
statistics, however they may contain distinct new information for
example about the oxygenation of tissues. It could be obtained
alongside any routine scan, e.g. FDG PET, using photons that are
currently wasted, so it is certainly worth further exploration.

\ack This work has been supported by the UK Research Councils
Basic Technology Programme.

\References

\item[]
 Charlton M and Humberston J W 2001 Positron Physics (Cambridge,
 UK: Cambridge Univ. Press)

\item[] Cooper A M, Laidlaw G J and Hogg B G 1967 Oxygen quenching
of positron lifetimes in liquids \textit{\JCP} \textbf{46}
2441-2442

\item[] De Benedetti S, and Siegel R 1954  The three-photon
annihilation of positrons and electrons \textit{Phys. Rev.} {\bf
94} 955-959

\item[] Chang T and Tang H 1985 Gamma-ray energy spectrum from
orthopositronium three-gamma decay \textit{Phys. Lett. B}
\textbf{157B} 357-360

\item[] Drezet A, Monnet O, Mont\'{e}mont G, Rustique J, Sanchez G
and Verger L 2004 CdZnTe detectors for the positron emission
tomographic imaging of small animals \textit{IEEE Nucl. Sci. Symp.
Conf. Record } R11-67

\item[] Hopkins B and Zerda T W 1990 Oxygen quenching of
positronium in silica gels \textit{\PL A} \textbf{45} 141-145

\item[] Kacperski K, Spyrou N M and Smith F A 2004 Three-gamma
annihilation imaging in positron emission tomography \textit{IEEE
Trans. Med. Im.} M9-451

\item[] Kacperski K and Spyrou N M 2004 Three-Gamma Annihilations
as a New Modality in PET \textit{IEEE Nucl. Sci. Symp. Conf.
Record }

\item[] Kakimoto M, Hyodo T and Chang T B 1990 Conversion of
ortho-positronium in low-density oxygen gas \textit{\jpb}
\textbf{23} 589-597

\item[] Klobuchar R L and Karol P J 1980 \textit{\JCP} \textbf{84}
483

\item[] Machulla H J 1999 Imaging of hypoxia. Tracer developments.
(Kluwer Academic Publishers)

\item[] Moses W W, Derenzo S E and Budinger T F 1994 PET detector
modules based on novel detector technologies  \textit{Nucl. Inst.
Meth. A} \textbf{353} 189-194

\item [] Moss C E, Ianakiev K D, Prettyman T H, Smith M K and
Sweet M R 2001 Multi-element, large-volume CdZnTe detectors
\textit{Nucl. Inst. Meth. A} \textbf{458} 455-60

\item [] Nemirovsky Y, Asa G, Gorelik J and Peyser A, 2001 Recent
progress in n-type CdZnTe arrays for gamma-ray spectroscopy
\textit{Nucl. Inst. Meth. A} \textbf{458} 325-333

\item [] Ore A and Powell J L 1949 Three-photon annihilation of an
electron-positron pair \textit{Phys. Rev.} \textbf{75} 1696-1699

\item[] Philips B F, Kroeger J D, Kurfess J D, Johnson W N, Wulf E
A and Novikova E I 2002 Small animal PET imaging with germanium
strip detectors \textit{IEEE Nucl. Sci. Symp. Conf. Rec.}
\textbf{3} 1438-1442

\item[] Redus R, Huber A, Pantazis J, Pantazis T, Takahashi T and
Woolf S 2004 Multielement CdTe stack detectors for gamma-ray
spectroscopy \textit{IEEE Trans. Nucl. Sci.} {\bf 51} 2386--2394

\item[] Scheiber C and Gaikos G C 2001 Medical applications of
CdTe and CdZnTe detectors \textit{Nucl.Inst. Meth. A} \textbf{458}
12-25

\item[] Schepp L A and Vardi Y 1982 Maximum likelihood
reconstruction for emission tomography \textit{IEEE Trans. Med.
Im.} {\bf MI-1} 113--122

\item[] Stickel J R and Cherry S R 2005 High-resolution PET
detector design: modelling components of intrinsic spatial
resolution \textit{\PMB} \textbf{50} 179-195

\item[] Verger L, Drezet A, Gros d'Aillon, Mestais C, Monnet O,
Mont\'{e}mont G, Dierre and Peyret O 2004 New perspectives in
gamma-ray imaging with CdZnTe/CdTe \textit{IEEE Nucl. Sci. Symp.
Conf. Record} JRM1-1

\item[]

\endrefs

\end{document}